# Effect of Gas Pressure and Molecular Structure in the Spatial-Mode Conversion of Terahertz Radiation from a Plasma Filament

Y.E. Geints[1], L.V. Seleznev[2], G.E. Rizaev[2], T.A. Dick[2], M.V. Levus[2]
[1] V.E. Zuev Institute of Atmospheric Optics, Russian Academy of Sciences, 1 Acad. Zuev Square, Tomsk 634055, Russia
[2] P.N. Lebedev Physical Institute, Russian Academy of Sciences, 53 Leninsky Prospekt, Moscow 119991, Russia

## Abstract

In this work, the generation of terahertz (THz) radiation under single-color filamentation of a femtosecond laser pulse in an optical cell with compressed gases (argon, nitrogen, $CO_2$) in broad pressure range from 0.001 to 25 atm is experimentally studied for the first time. A nonmonotonic dependence of the integral energy of the THz signal on the gas pressure is revealed, which can be attributed to the competition between an increase in the plasma density and the development of plasma self-absorption processes, multiple filamentation of the femtosecond pump, and collision-induced absorption in the neutral medium. We show that the molecular structure of the gas critically determines the pressure optimum of generation, shifting the maximum of the THz energy yield to the region of elevated pressures (10–15 atm) for monatomic argon and to the region of subatmospheric pressures (~0.1 atm) for molecular carbon dioxide due to strong vibrational-rotational dissipation. On the basis of numerical 3D modeling, it is shown that the cylindrical metal waveguide-insert used for signal collection operates in a strongly multimode regime and substantially transforms the angular spectrum of the THz source. We demonstrate that the conical structure of the THz radiation recorded in the experiment is a consequence of intermode interference and diffraction at the output end face of the waveguide, which transforms the true parameters of THz generation in the plasma channel.

## 1. Introduction

The generation of terahertz radiation under femtosecond laser filamentation in gases represents one of the most actively developing fields of modern nonlinear optics and plasma physics [1, 2]. Interest in this area is driven both by the fundamental significance of the processes accompanying the propagation of ultrashort and high-intensive laser pulses in a medium and by the prospects for creating compact THz radiation sources for spectroscopy, substance diagnostics, and applied problems in the field of communications and sensing [3, 4]. Under conditions of filamentation in a gaseous medium, the generation of THz radiation is caused by several physical mechanisms, each of which dominates in a certain range of radiation and medium parameters.

In the strong optical field characteristic of femtosecond laser pulses, multiphoton/tunnel ionization of gas atoms and molecules dominates, during which active generation of a underdense plasma occurs in the radiation channel, in which the excited transient photocurrents serve as the main source of THz radiation under filamentation in the intensity clamping regime [5]. Numerical modeling of femtosecond filamentation shows that photocurrents dominate over alternative THz generation mechanisms, including Kerr effects, transition Cherenkov radiation [6], and longitudinal plasma sources [7]. This is explained by the fact that in the intensity clamping regime of a filament, it is the asymmetric acceleration of free electrons in the pulse field that efficiently converts optical energy into low-frequency submillimeter radiation. However, it is important to note that under other regimes of laser pulse filamentation, the relative contribution of the photocurrent can change substantially, and its dominance can become non-absolute, in particular, a dependence on pressure and beam geometry can be revealed [7-9]. Obviously, the efficiency of THz generation mechanisms is determined by the parameters of the laser pulse, the gas composition, its pressure, as well as the spatiotemporal structure of the forming filament. Of particular interest here is the study of the influence of gas pressure, since it simultaneously changes the density of the neutral medium, the ionization conditions, the filament length, the collision frequency of charged particles and, as a consequence, the balance between generation and absorption of THz radiation.

Recently, it is established that the energy of THz radiation grows quasi-linearly with the energy of the pumping laser pulse under both single-color and two-color filamentation [10-12]. This indicates the scalability of the process of laser-induced low-frequency signal generation. However, the linearity of the dependences can be violated at very high pulse energies, when multiple filamentation occurs in the medium. For example, the use of a noisy super-Gaussian laser pulse profile with multiple intensity spikes leads to energy distribution among several emerging filaments, which reduces the total plasma response and, consequently, the THz yield [5]. Thus, maintaining a single filament is a critical condition for achieving maximum conversion efficiency, which, as will be seen below, can be implemented, for example, using geometric focusing of radiation and waveguide THz structures.

The present work considers experimental results on the generation of submillimeter electromagnetic radiation in the spectral range of 0.1–10 THz during laser filamentation in three different gases (nitrogen, argon, $CO_2$) at variable pressure from vacuum values to 25 atm. These data are important for elucidating the potential of gas pressure for obtaining a new type of radiation by laser filamentation. The novelty of this study consists in the use of elevated-pressure gases, which were previously considered inefficient for THz generation. In addition, an essential feature of the experimental scheme used is the presence of a waveguide channel for delivering radiation to the detector by means of a metal cylindrical insert. This circumstance is fundamentally important for the interpretation of the results, since the recorded THz angular pattern is not a direct representation of the initial spatial-angular spectrum formed in the plasma channel, but represents the result of its transformation by the waveguide structure. Therefore, in analyzing the experimental results, the study takes into account not only the mechanisms of THz radiation generation, but also mode filtering, interference, and the spectral-angular transfer function of the cylindrical waveguide.

Thus, the aim of this work is to study the dependence of THz parameters on pressure and gas type, as well as to identify the conditions under which maximum generation efficiency is achieved. Of particular interest is the comparison of the central and peripheral components of the obtained angular distribution with the integral energy dependence of the signal, which allows one to judge the competition between generation processes, plasma self-absorption, and collision-induced absorption (CIA) of molecules, as well as waveguide transformation of the radiation. One of the main results of this paper is the demonstration of recording THz radiation from an optical cell under filamentation of a femtosecond pulse in elevated-pressure gases, which, to our knowledge, is performed experimentally for the first time. Equally important is the establishment of the dependence of the THz radiation energy on the type and pressure of the gas in the cell. The obtained results are important for understanding the physics of THz generation in compressed gases and for optimizing the parameters of sources based on laser filamentation.

## 2. Experimental Results and Discussion

In the experiments, we use pulses of a titanium-sapphire laser system with a central wavelength of 750 nm at a duration of 90 fs. The laser beam diameter at the 1/e level is 8 mm. To generate THz radiation at different pressures in different gases, we use a two-meter stainless-steel gas cell (a seamless steel pipe with an inner diameter of 50 mm) with 8-mm-thick sapphire windows to reduce energy absorption in the THz range. The laser beam is focused into the cell by a lens with a focal length of 1 m so that the beam intensity at the input/output windows of the cell is minimal. In addition, inside the cylindrical cell, closer to the output window, a 1-m-long brass insert with an inner diameter of 14 mm and a wall thickness of 1 mm is added (see Fig. 1a and b). In fact, such an insert represents a metal waveguide with a high reflection coefficient from the walls, which makes it possible to intercept and directionally extract almost all emerging THz radiation from the cell into a size comparable to the input aperture of a vacuum bolometer (10 mm) with minimal losses.

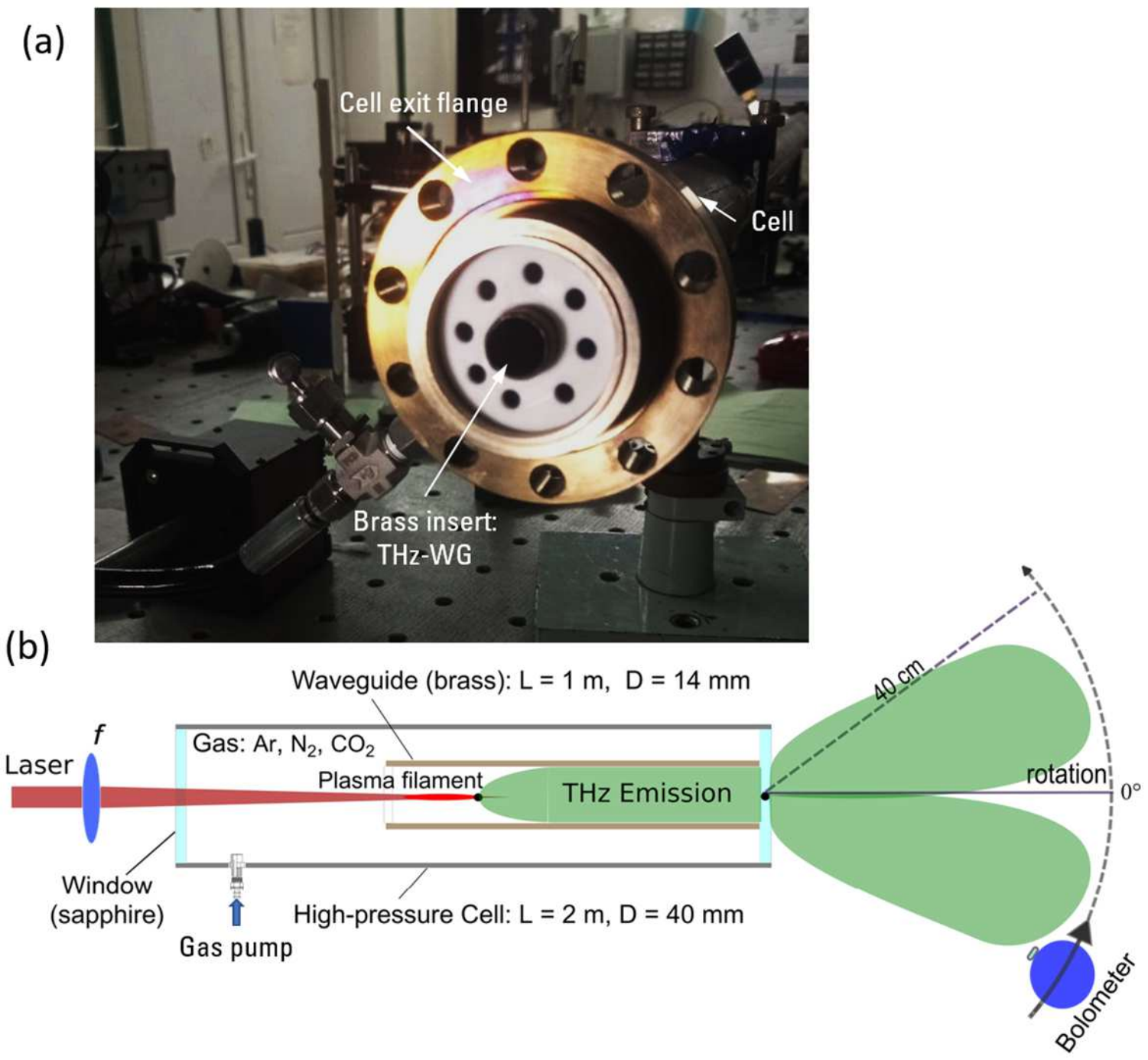

**Fig. 1** Experimental setup and principal scheme of terahertz signal detection. **(a)** Photograph of the output flange of a 2-m stainless-steel high-pressure optical cell with a coaxially installed 1-m brass cylindrical insert with an inner diameter of 14 mm, which performs the function of a THz waveguide–optical cell with a waveguide insert. **(b)** Principal scheme of experiments on THz radiation generation by single-color laser filamentation of fs pulses of a Ti:Sapphire laser in a gas cell under various pressures.

For THz radiation detection, a cryogenic bolometer operating in the range from 0.1 to 6 THz is used. The bolometer is located at a distance of 40 cm from the output aperture of the waveguide on a movable console that moves in the horizontal plane around the output aperture of the metal waveguide to obtain angular distributions of THz radiation. We note that previously a similar modification of the traditional scheme of THz generation from a laser filament is used by us for atmospheric air [13]. THz radiation is recorded without the use of spectral filters; therefore, all generated frequencies of the long-wavelength signal enter the receiver. The angular dependences of THz radiation at different gas pressures are shown in three panels of Fig. 2 for argon, nitrogen, and carbon dioxide.

From the analysis of the data obtained, two characteristic angular components of THz radiation from the filament region can be distinguished. At all studied pressures, pronounced central maxima are observed at small angles (5–7 degrees) to the propagation direction of the exciting laser beam, which in the literature is traditionally associated both with the transition Cherenkov emission mechanism of a plasma source moving at a superluminal velocity [6], and with the superposition in the far zone of the radiation of a travelling-wave source formed by THz radiation of longitudinal and transverse photocurrents arising from the rapid ionization of the gas and subsequent plasma relaxation [14, 15].

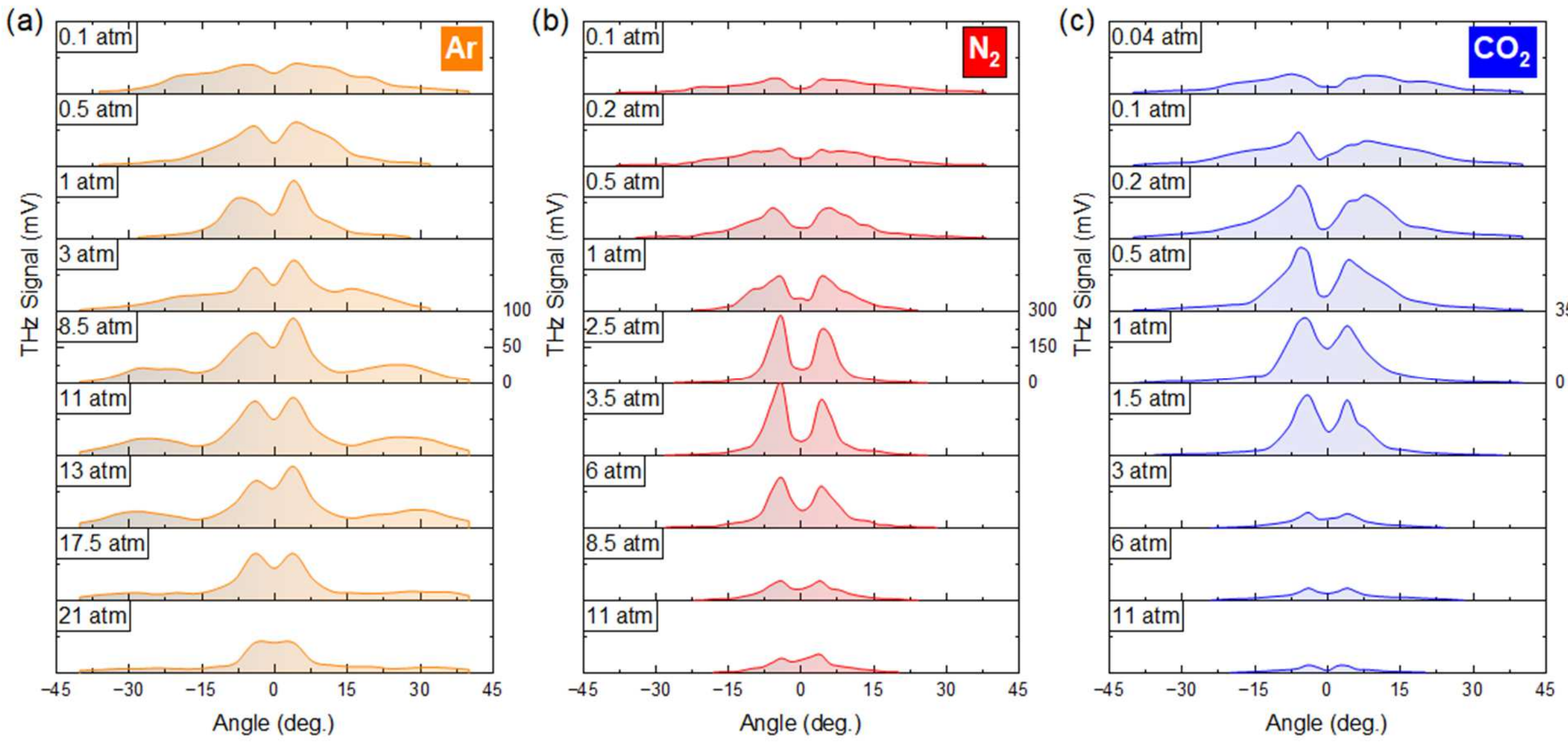

**Fig. 2** Experimental data on the angular distribution of THz radiation at the output of the waveguide of a cell with gases of different pressure (indicated in the legend) for **(a)** argon, **(b)** nitrogen, and **(c)** carbon dioxide. The graphs demonstrate a rearrangement of the angular spectrum from a weakly pronounced distribution at deep vacuum to a strongly differentiated two-component structure with a central maximum and side lobes of conical emission, the degree of symmetry, broadening, and intensity of which are determined by the gas type and the effective nonlinearity of the medium.

At the same time, as is seen in the distributions of Fig. 2, with increasing pressure from tenths of an atmosphere to several tens of atmospheres, side lobes, or “wings,” clearly appear in the polar diagrams, the angular position of which varies depending on the gas type. The appearance of these peripheral maxima can be associated with an increase in the contribution of conical emission caused by the nonstationary photocurrent of free electrons of the plasma filament channel [15, 16]. Under conditions of high gas density, the third-order nonlinear susceptibility increases sharply, and self-focusing of the laser beam leads to the formation of an extended filament, inside which phase-matching conditions are satisfied for the generation of terahertz waves at nonzero angles to the axis. Meanwhile, the degree of manifestation of the side lobes depends substantially on the molecular structure of the gas. Indeed, for the argon atom, which has no rotational degrees of freedom, the angular THz wings remain sufficiently clear and symmetric, whereas in nitrogen they are somewhat blurred due to energy redistribution over vibrational-rotational levels, and in carbon dioxide the greatest width and asymmetry of the angular spectrum is observed, which is explained by the anomalously high cubic nonlinearity coefficient [17] and the simultaneous enhancement of dissipative processes associated with electron-molecular collisions.

The integral characteristic of THz radiation from the filament region, obtained by summing the signal over all reception angles with allowance for cylindrical symmetry, is presented in Figs. 3a and b. This integral parameter gives the dependence of the total terahertz radiation energy on pressure and gas type. Thus, from the analysis of these figures, a nonmonotonic form of all curves follows, when at first an increase in THz energy is observed with increasing pressure up to a certain optimal value, after which a decline follows. In the initial section, the growth of the number of molecules per unit volume (growth of gas pressure) leads to an increase in the plasma density and the length of the forming light filament, which proportionally increases the effective volume of THz generation due to both the Cherenkov and ponderomotive mechanisms. However, with further gas compression, when the pressure exceeds 10–15 atm (depending on the gas type and pump energy), THz self-absorption and optical pump fragmentation processes begin to play a dominant role, leading to a decrease in THz radiation energy.

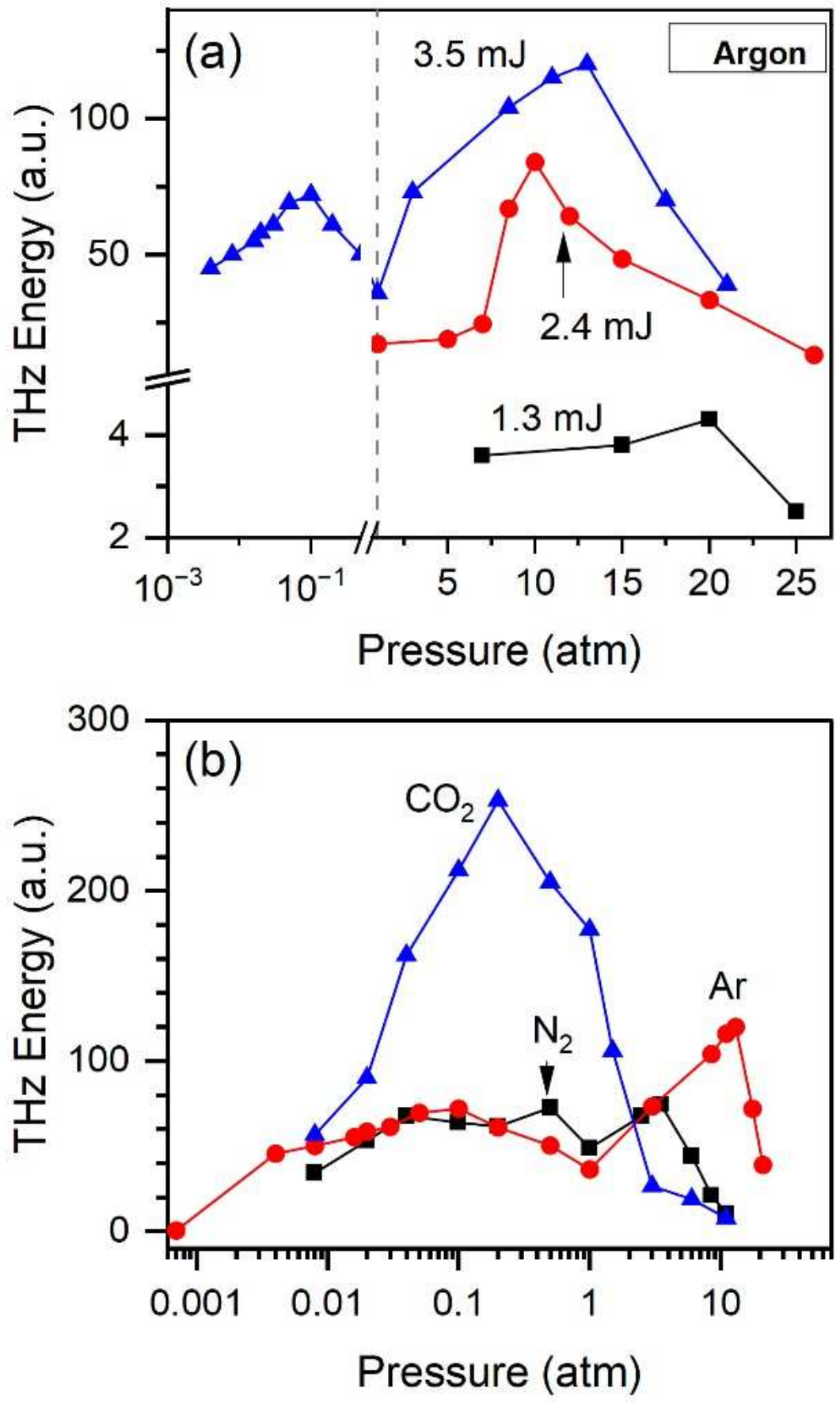

**Fig. 3** Pressure dependences of the THz radiation integrated over angles from the optical cell. **(a)** THz energy for argon at different optical pump energies. **(b)** Comparison of the integrated THz signal yield for the three studied gases at a fixed femtosecond pulse energy $E$ = 3.5 mJ.

Indeed, a high concentration of free electrons in the plasma leads to a significant increase in the frequency of their collisions with neutral particles, which causes strong attenuation of terahertz waves directly inside the plasma channel created by the propagating laser pulse. This is due to strong inverse bremsstrahlung absorption of THz waves by free electrons [6, 16]. In addition, at high gas density, increased absorption of electromagnetic radiation also occurs inside the cell volume in the neutral gas itself, which occurs due to the so-called collision-induced absorption (CIA) [18]. This effect arises only for actively colliding particles (molecules of compressed gas), each of which individually in this spectral range may not absorb. At the same time, during the microscopic act of collision, the symmetry of the system is violated and a transient (induced) dipole moment arises, which interacts with the photon of the incident radiation. The CIA effect is specific only to molecular gases and is completely absent in argon, and in the "heavy" $CO_2$ molecule such collision-induced absorption represents a smooth continuum extending over hundreds of wavenumbers from fractions to tens of THz [19]. At the maximum of the absorption band with a center near 2 THz, the induced absorption coefficient of $CO_2$ can reach values of the order of several $cm^{-1}$ at a gas pressure of 20 atm [18], which leads to complete attenuation of the THz signal already at a distance of a few centimeters. Another reason for the decrease in radiation in the terahertz range at sufficiently high gas pressures can be a change in the filamentation regime of the femtosecond pulse itself. Under such conditions, multiple filamentation of the laser pulse arises, which fragments the initial beam, lowers the local intensity, and, consequently, the density of free electrons in each individual filament, and shortens the effective length of the plasma channel, thereby reducing the output of the microwave signal [11, 20].

At a fixed femtosecond pulse energy of 3.5 mJ, as shown in Fig. 3b, attention is drawn to the existence of an optimal pressure at which the maximum THz radiation yield is achieved. However, the values of this maximum, its amplitude, and the steepness of the decline differ strongly. The highest integral yield is observed for carbon dioxide, whereas argon is inferior by approximately a factor of 1.5–2, and nitrogen gives the lowest low-frequency signal. This difference can be due to a combination of a number of factors.

Argon, being a monatomic gas, practically does not lose energy to vibrational excitation, which ensures high efficiency of conversion of optical radiation into low-frequency microwave. Nitrogen spends part of its energy on excitation of vibrational levels of molecules, which affects the decrease in effective nonlinear polarization and the absence of a distinct extremum in the pressure dependence of low-frequency radiation. The $CO_2$ molecule has the richest system of vibrational levels, which are actively excited in an intense laser field. At the same time, it is precisely this feature of the structure of the carbon dioxide molecule that leads to an increased Raman response and a significant rise in the effective nonlinear refractive index especially at high pressures [17], which increases the total density of generated plasma in the laser beam channel and the efficiency of conversion into a low-frequency signal. On the other hand, strong low-frequency deformation modes are present in the $CO_2$ molecule, which enhance rotational relaxations and efficiently dissipate energy in the terahertz range with increasing probability of molecular collisions via CIA mechanism. As a result, even at relatively low pressures, losses to plasma absorption and energy transitions into vibrational-rotational modes exceed the growth of generation processes. Therefore, the THz maximum in $CO_2$ is achieved at a reduced gas density, approximately at 0.2 atm, and a further increase in pressure only increases dissipation without leading to a significant increase in the output signal.

It is important that a joint consideration of the angular distributions and integral energy dependences allows one to conclude that the maximum generation efficiency is achieved either at subatmospheric or at moderately high pressures, when conical lobes in the angular diagram only begin to manifest themselves but substantial plasma and intrinsic absorption does not yet occur. The most promising medium for THz radiation generation at elevated pressure (~10 atm) is argon, and the use of carbon dioxide is expedient only at moderately reduced pressure (~0.1 atm) due to strong dissipation of microwave energy.

## 3. Influence of the waveguide collecting system on the THz angular distribution

We recall that for collecting the low-frequency THz radiation generated by the plasma of a femtosecond filament, in our experiments we use a cylindrical metal waveguide placed inside the gas cell immediately after the geometric focus of the beam. Thus, the waveguide covers the plasma region, concentrates and directs the emerging THz waves to the exit from the cell into a narrow aperture of about 14 mm. This allows us to substantially amplify the amplitude of the low-frequency signal received by the bolometer and to measure the angular diagram more accurately.

However, it should be taken into account that the measured angular distributions are formed not directly in free space, but after propagation of THz radiation through a sufficiently long metal cylindrical insert. Therefore, the recorded diagram is the result of transformation of the initial spatial-angular spectrum of the THz source by the waveguide structure [13]. The metal walls of the insert ensure efficient collection of radiation propagating at nonzero angles to the axis, but at the same time lead to excitation of a large set of eigenmodes of the waveguide of various polarizations, differing in propagation coefficients, losses, and phase incursion. Thus, in the typical detection range of low-frequency radiation under single-color filamentation of 0.1–1 THz, a metal pipe with a diameter of 14 mm is a substantially multimode cylindrical waveguide, i.e., the cutoff frequency of such a waveguide is substantially lower than the studied THz range. At a frequency of 0.1 THz, the first modes correspond to effective angles of the order of 7–20°, whereas at 1 THz the same modes correspond to angles of less than 2°. At the same time, higher-order modes allow the transfer of significantly wider angular components; therefore, the waveguide does not limit THz radiation by a simple geometric angle but performs its spatial-mode transformation. As a consequence, the position and relative intensity of the side lobes at the waveguide output (see, Fig. 2) cannot be considered as a direct quantitative representation of the generation angles in the plasma channel. In particular, mode filtering and interference can lead to smoothing, enhancement, or redistribution of energy between the central and peripheral regions of the diagram, which means that the development of the conical structure observed in the experiments with increasing gas pressure should be interpreted primarily as a qualitative sign of the appearance of nonzero angular components of THz radiation, whereas their absolute angular position requires separate consideration of the transfer function of the waveguide insert. In particular, the angle of the maximum observed by the bolometer can differ from the angle of conical emission directly in the plasma channel.

In order to demonstrate the effect of the waveguide, we perform numerical modeling of the propagation of radiation in a cylindrical copper waveguide at two THz frequencies characteristic of the microwave spectrum under single-color filamentation [21]. The computer model is built using the COMSOL Multiphysics package in a 3D-geometry (in fact, 2D-axisymmetric). A 10-cm section of a hollow cylindrical brass pipe with a wall thickness of 1 mm and a diameter of 14 mm acts as the waveguide. The plasma source of THz waves is modeled as an elongated ellipsoid located on the waveguide axis with a length of 1 cm and a transverse diameter of 100 μm, which resembles the real plasma channel in the focused geometry of single-color filamentation. To specify the plasma photocurrent $\mathbf{J}_e(r,z)$ generating low-frequency radiation, the "External Current Density" node of the "Wave Optics" module of COMSOL is used, in which the density of external electric currents is specified in the following form:

$$\mathbf{J}_e(r,z) = J(r)F(z)\exp(-jk_{\mathrm{THz}}z) \tag{1}$$

Here, following the formalism of [6], the dependence of the plasma photocurrent density on the longitudinal and transverse coordinates is represented in factorized form.

In Eq. (1), $J(r)$ takes into account the radial character of the low-frequency plasma current, since it is assumed that the THz field in the single-color regime is generated due to ponderomotive expulsion of electrons from the filament axis across the beam. Since the profile of the laser beam itself in the filament is close to Gaussian, the ponderomotive force, proportional to the intensity gradient, is mathematically described by the spatial derivative of the Gaussian function as follows:

$$J(r) \propto (r/r_F)\exp\left\{-(r/r_F)^2\right\}, \tag{2}$$

where $r_F$ is the effective radius of the plasma channel.

In a realistic femtosecond filament, the electron density grows from zero at the beginning of the channel, reaches a peak of ~$10^{15}$–$10^{17}$ cm$^{-3}$ (for air) in the waist region, and smoothly decreases at the exit due to plasma recombination and defocusing of the laser pulse. Mathematically, this means that the photocurrent density becomes a function of the z coordinate and can be approximated by a Gaussian:

$$F(z) \propto \exp\left\{-\left[(z-z_0)/L_F\right]^2\right\}, \tag{3}$$

where $L_F$ is the effective longitudinal size of the plasma channel with the center at $z_0$.

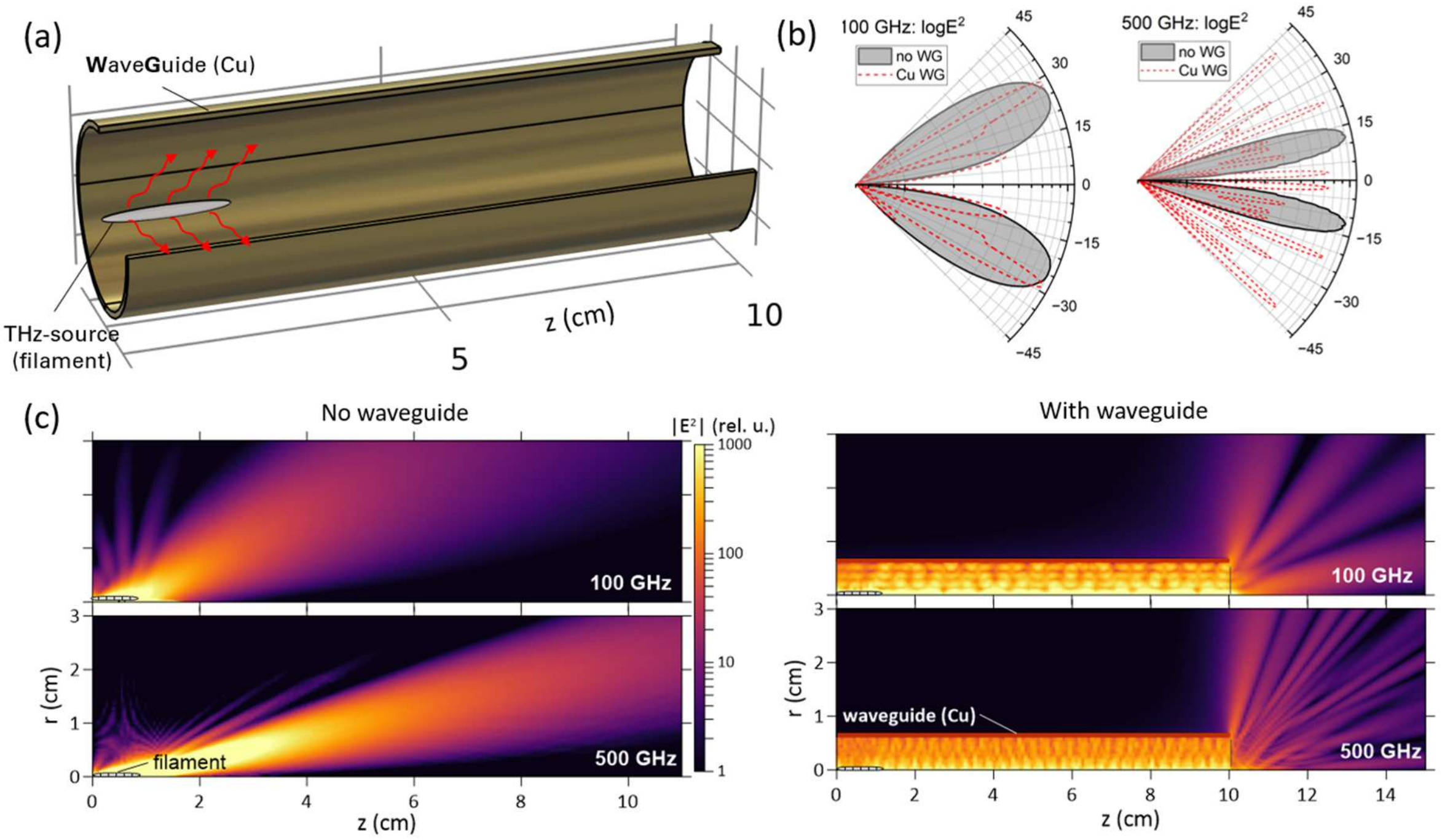

**Fig. 4** Results of numerical 3D simulation of the spatial-mode transformation of THz radiation in the waveguide tract by the finite element method in COMSOL Multiphysics. **(a)** Three-dimensional geometry of the computational domain, including an extended axisymmetric ellipsoidal plasma source with a specified ponderomotive radial photocurrent and a section of a brass cylindrical pipe with a diameter of 14 mm. **(b)** Far-field polar radiation patterns and **(c)** two-dimensional spatial distributions of the relative intensity of the THz field at frequencies of 100 GHz and 500 GHz, comparing free propagation of waves with their transportation inside the waveguide.

Finally, the last factor in Eq. (1) takes into account the phase delay of the THz wave with wavenumber $k_{\mathrm{THz}}$ emitted by the filament at the point with coordinate $z$. In addition, since the plasma possesses strong absorption in the THz range, this effect of energy attenuation inside the plasma channel itself is modeled as the appearance of an imaginary part of the wave vector. This leads to the addition in Eq. (1) of an exponential factor $\exp(-\alpha_{\mathrm{THz}} z)$, where the plasma absorption coefficient $\alpha_{\mathrm{THz}}$ is calculated using the Drude model with allowance for the change in plasma electron density according to dependence similar to Eq. (3). The entire computational model is conventionally enclosed in a system of perfectly matched layers (PML) preventing parasitic wave reflections from the domain boundaries. For discretization of the problem, a tetrahedral mesh with a maximum edge length $l = 1/k_{THz}$ is used.

The results of numerical simulation of the angular distribution of THz radiation from the filament after propagation in the waveguide are presented in Fig. 4. A comparison of the two-dimensional spatial distributions of the field intensity (Fig. 4c) shows that under free-space conditions (without the waveguide), classical conical emission is formed, the parameters of which demonstrate a pronounced spectral dependence. For the lower frequency of 100 GHz, the wavefront is characterized by a broad angular divergence with an angle of ~25°, whereas at 500 GHz the radiation becomes noticeably more narrowly directed and pressed to the optical axis (~13°), which is explained by the improvement of phase-matching conditions along the extended plasma channel for shorter wavelengths. When the source is placed inside the metal insert, terahertz waves undergo multiple reflections from the conducting walls, forming a characteristic interference pattern with alternating antinodes and nodes of intensity in the pipe volume. Since the inner diameter of the waveguide is many times greater than the wavelengths of the studied range, this structure operates in a strongly multimode regime, capturing and transporting a significant part of the generated energy, which at the output of the end face experiences diffraction at the circular aperture.

The influence of waveguide filtering is clearly illustrated by the far-field radiation patterns in Fig. 4b. At a frequency of 100 GHz, the initial two-lobe radiation profile with strictly zero intensity on the beam axis, caused by the radial character of the photocurrent, after passing through the waveguide generally retains its geometry but acquires additional amplitude modulation and forms additional lobes at an angle of ~10°. On the contrary, at a frequency of 500 GHz, the initial narrow radiation cone undergoes a radical rearrangement and at the waveguide output breaks up into a fan of numerous narrow diffraction lobes propagating at substantially higher angles compared to free space.

We note that in our numerical modeling, a short 10-cm waveguide is considered, whereas in reality its length is about one meter. It is clear that an increase in the waveguide length mainly transforms the phase and amplitude profile of the transfer function due to dispersive spreading and spatiotemporal separation of modes. Since each mode possesses its own frequency-dependent group velocity, lengthening of the waveguide leads to an increase in intermode dispersion. This leads to the appearance of deep high-frequency oscillations in the transfer function of the waveguide structure, caused by phase mismatch and destructive interference of modes of different orders at the output end face of the channel [22]. In addition, the waveguide length also affects the exponential attenuation of the signal through the individual mode loss coefficients. At the same time, since energy losses on the metal walls for higher-order modes are substantially higher, an increase in the waveguide length is analogous to additional mode filtering, which removes components with large tilt angles from the transfer function and leaves at the output predominantly the lowest modes characterized by the smallest ohmic attenuation [23]. Thus, the simulation results clearly show that the angular distribution recorded in the experiment is not a direct representation of the initial generation angles in the plasma channel but represents a complex superposition of excited waveguide modes, which requires mandatory consideration in the quantitative interpretation of experimental data.

## 4. Conclusion

In conclusion, during our experimental study of THz generation under femtosecond laser filamentation in rarefied and compressed gases we found that the integral energy of the THz signal depends nonmonotonically on the medium pressure, which can be associated with the competition between the growth of plasma density and the development of self-absorption processes, pulse fragmentation, as well as collision-induced absorption in the neutral gas. The determining influence of the molecular structure of the gas on the pressure optimum of THz signal generation is revealed. In argon, the maximum THz energy is achieved at pressures of 10–15 atm, whereas in carbon dioxide, due to strong vibrational-rotational dissipation, the highest yield shifts to the subatmospheric region of about 0.1 atm. Using numerical 3D simulation, it is shown that the cylindrical metal insert used for signal collection operates in a strongly multimode regime and substantially transforms the angular spectrum of the THz source. It is demonstrated that the conical structure of THz radiation recorded in the experiment is a consequence of intermode interference and diffraction at the output end face of the waveguide, which indicates the need to take into account its transfer function when determining the true parameters of generation in the plasma channel.

The obtained results are important for understanding the physics of THz generation in compressed gases and for optimizing the parameters of sources based on laser filamentation. The revealed dependence of the pressure optimum on the molecular structure of the gas indicates the need for a targeted choice of the gas medium and its pressure for specific applied problems. In particular, argon at elevated pressures is the most promising medium for creating compact high-power THz sources, whereas carbon dioxide is expedient only at moderately reduced pressures due to strong collision-induced absorption. Further development of this work is associated with the reconstruction of the true spatial-angular spectrum of the source based on the measured waveguide transfer function, as well as with the study of the influence of the waveguide length and diameter on the efficiency of THz radiation collection.

**Funding.** The experimental study of terahertz radiation was supported by the Russian Science Foundation (grant No. 24-12-00056). The numerical 3D modeling was carried out within the framework of the state assignment of Ministry of Science and Higher Education of Russia/IAO SB RAS.

**Conflicts of Interest.** The authors declare that they have no competing financial interests or personal relationships that could have appeared to influence the work reported in this paper.

**Data Availability Statement**

The datasets generated during and/or analyzed during the current study are available from the corresponding author on reasonable request.

**Authors' Contributions.** All authors contributed to the study conception and design. Material preparation, data collection, and experiments were performed by L.V. Seleznev and G.E. Rizaev. Numerical 3D simulations and theoretical analysis were conducted by Y.E. Geints. The first draft of the manuscript was written by Y.E. Geints, and all authors commented on previous versions of the manuscript. All authors read and approved the final manuscript.